\documentclass{rspublic}
\usepackage{graphicx}
\usepackage[numbers, square, sort&compress]{natbib}

\begin{document}

\title[Joining Bayesian and Frequentist Methods]{Joining Forces of Bayesian and Frequentist Methodology: A Study for Inference in the Presence of Non-Identifiability}

\author[A. Raue and others]{Andreas Raue$^{1}$, Clemens Kreutz$^{1}$, \\Fabian Joachim Theis$^2$ and Jens Timmer$^{1,3}$}

\affiliation{$^1$ Institute for Physics, University of Freiburg, Germany\\
$^2$ Helmholtz Zentrum Munich and Departement of Mathematics, Technical University of Munich, Germany\\
$^3$ BIOSS Centre for Biological Signalling Studies and Freiburg Institute for Advanced Studies (FRIAS), Freiburg, Germany; Department of Clinical and Experimental Medicine, Link\"oping University, Sweden}

\label{firstpage}

\maketitle

\begin{abstract}{identifiability, profile likelihood, Bayesian Markov chain Monte Carlo sampling, posterior propriety, propagation of uncertainty, prediction uncertainty}
Increasingly complex applications involve large datasets in combination with non-linear and high dimensional mathematical models. In this context, statistical inference is a challenging issue that calls for pragmatic approaches that take advantage of both Bayesian and frequentist methods. The elegance of Bayesian methodology is founded in the propagation of information content provided by experimental data and prior assumptions to the posterior probability distribution of model predictions. However, for complex applications experimental data and prior assumptions potentially constrain the posterior probability distribution insufficiently. In these situations Bayesian Markov chain Monte Carlo sampling can be infeasible. From a frequentist point of view insufficient experimental data and prior assumptions can be interpreted as non-identifiability. The profile likelihood approach offers to detect and to resolve non-identifiability by experimental design iteratively. Therefore, it allows one to better constrain the posterior probability distribution until Markov chain Monte Carlo sampling can be used securely. Using an application from cell biology we compare both methods and show that a successive application of both methods facilitates a realistic assessment of uncertainty in model predictions.
\end{abstract}

\section{Introduction}
In many scientific situations mathematical models are used to predict properties of a system under study using explicitly formulated assumptions and hypotheses. This is especially important for complex applications that do not allow for an intuitive understanding of the processes involved. Applications range from analyses in particle physics \citep{Feroz:2011fk} or in climate research \citep{Smith:2009fk, Moncho:2012fk} to quantifying dynamical processes in cell biology \citep{Bachmann:2011fk}. Often, these mathematical models contain parameters that are unknown or known only with large uncertainty. For example in bio-chemical models that are utilised in cell biology, parameters such as reaction rate constants, amount of molecular compounds, detection sensitivities or measurement backgrounds are often unknown. Before a model can be used for prediction reliably, the unknown parameters have to be estimated by comparing model output to experimental data. 

For a realistic assessment of the accuracy of model predictions, it is important that uncertainties in the experimental data and in prior assumptions are propagated correctly via the parameters to the desired predictions. Bayesian Markov chain Monte Carlo (MCMC) sampling facilitates this propagation of uncertainties by sampling from the posterior probability distribution \cite{Robert:2004fk}. However, if experimental data is limited and the mathematical models are non-linear and contain many unknown parameters the posterior probability distribution can be insufficiently constrained. For such insufficiently constrained posterior the probability mass can be distributed widely in a high dimensional parameter space. Consequently, MCMC sampling can quickly become infeasible.

Alternatively, one can resort to frequentist methods in this situation. Here, insufficient experimental data and prior assumptions can be interpreted as non-identifiability of the model parameters \cite{Walter:1997uq}. We applied a generic approach that uses the profile likelihood to detect both structural and practical non-identifiability \citep{Raue:2009ec}. Furthermore, this approach allows one to design new experiments that resolve non-identifiability. Therefore, it is beneficial to further constrain the posterior probability distribution until MCMC sampling can be applied reliably and efficiently. 

We compared the results of both MCMC sampling and profile likelihood methods. In the absence of non-identifiability the results of both methods are in good agreement. However, in the presence of non-identifiability their results can be substantially different. Our results imply that MCMC sampling in the presence of non-identifiability can be misleading. Therefore, we suggest a successive application of both methods that ensures a realistic assessment of uncertainty in model predictions.

\subsection{Frequentist methods}
\begin{figure} 
\begin{center}
\includegraphics[width=\textwidth]{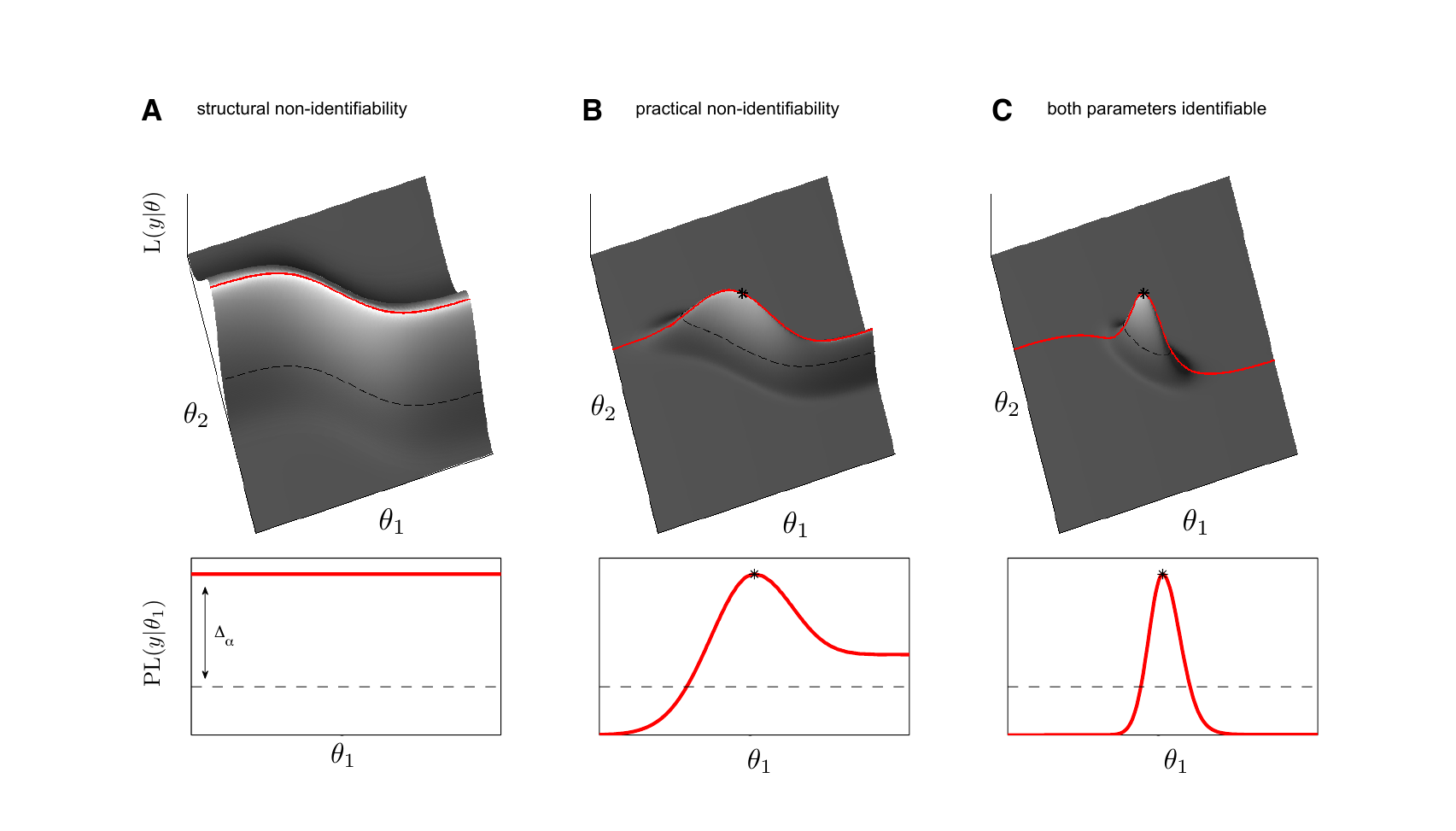}
\end{center}
\caption{Identifiability analysis using likelihood profiles: The upper panels show as illustration the shape of the likelihood $L(y|\theta)$ in two dimensions for three typical cases. The traces of the profiles in parameter space are indicated as red lines. 
The lower panels shows the respective profile likelihood $PL(y|\theta_1)$ for the dimension of parameter $\theta_1$ as red lines. In all panels the asterisks denote the MLE in cases where a unique solution exists and the dashed lines denote the threshold $\Delta_\alpha$ that yields a likelihood based confidence region \citep{Box:1973kx, Press:1990rw}. Three typical cases can arise: (A) A flat profile indicates structural non-iden\-tifiability. In this case, no unique solution for MLE exist. (B) A profile that decreases but tails out to a plateau to one or both sides indicates practical non-identifiability. (C) A profile that tails out to zero on both sides quickly enough, i.e.~at least exponentially fast, indicates structural and practical identifiability. The confidence interval of $\theta_1$, for (A) is infinite, for (B) has only a lower bound, for (C) has a finite range.}
\label{ples}
\end{figure}

Maximum likelihood estimation of model parameters is a theoretically well developed area \citep{Fisher:1922le}. Based on an assumption on the distribution of the measurement noise, the likelihood function $L(y|\theta)$ of the data $y$ given the parameters $\theta$ describes the agreement of model output and experimental data. In case of normally distributed measurement noise $\epsilon \sim \mathrm{N}(0,\sigma^2)$ the likelihood reads as 
\begin{equation}
	L(y|\theta) = \prod_{k=1}^m \prod_{l=1}^{d_k}\, \frac{1}{\sqrt{2\pi\sigma_{kl}^2}}\, \exp \left(- \frac{1}{2} \left(\frac{y_{kl} - y_{k}(t_{l}, \theta)}{\sigma_{kl}}\right)^2 \right) \label{llhoodfun}
\end{equation}
where $m$ model outputs $y_{k}(t_{l}, \theta)$ and $d_k$ data instances for each model output such as time points $t_l$ can be considered. The maximum of $L(y|\theta)$, i.e.~the best fit of the model to the data, provides a point estimate $\hat \theta$ of the parameters. This maximum likelihood estimate (MLE) can be calculated for non-linear models by numerical optimisation methods, see e.g.~the trust region algorithm in \citep{Coleman:1996fk} and for a general introduction in \citep{Press:1990rw, Seber:2003kq}. The uncertainty of the estimate $\hat \theta$ is buried in the shape of the likelihood function. Figure~\ref{ples} shows an illustration of the likelihood for three typical cases. 

If the amount of model quantities that can be accessed by experiments is limited, a subset of parameters can be structurally non-identifiable. This indicates that the parametrisation of the model is such that two or more parameters can compensate their effects and yield exactly the same model outputs $y_{k}(t_{l}, \theta)$. This in turn results in a constant likelihood value on a sub-manifold, see figure~\ref{ples}A. Consequently, the MLE for the parameters cannot be determined uniquely. The parameter relations that cause the structural non-identifiability are akin to gauge invariances in physical theories. However, for complex models it can be difficult to detect structural non-identifiability. For example, if models can only be evaluated by numerical simulation, such as in the case of detector models used in particle physics \cite{Allison:1992fk} or dynamical models that are used in cell biology \citep{Bachmann:2011fk}, structural non-identifiability cannot be detected directly. In the latter case, methods for \emph{a priori} structural non-identifiability analysis exist that analyse the structure of the ordinary differential equations (ODE) without having an analytical solution. A comparison of these methods can be found in \cite{Chis:2011fk}. However, \emph{a priori} methods are often limited to linear ODE systems or are impractical for models containing many parameters \cite{Raue:2010ys}. Arguably one of the the most practical \emph{a priori} methods that has also been proved to work for larger models applies a probabilistic algorithm \cite{Sedoglavic:2002tp,Karlsson:2012fk}.

In addition to structural non-identifiability, model parameter can be practically non-identifiable \citep{Raue:2009ec}. This type of non-identifiability arises if the amount and quality of experimental data is limited. It cannot be detected by \emph{a priori} methods. However, practical non-identifiabi\-li\-ty is of equal importance. A generic approach that allows one to detect both structural and practical non-identifiabi\-li\-ty at the same time uses the concept of the profile likelihood \citep{Raue:2009ec, Murphy:2000rp}. The profile likelihood $PL$ can be calculated for each parameter $\theta_i$ individually by
\begin{equation}
	PL(y|\theta_i) = \max_{\theta_{j\not=i}}[L(y|\theta)]. \label{profilellh}
\end{equation}
The equation indicates that for each value of $\theta_i$ all of the remaining parameters $\theta_j$ are re-optimised, see figure~\ref{ples} for illustration. The profiles $PL(y|\theta_i)$ break down the uncertainty contained in the high-dimensional likelihood $L(y|\theta)$ to a footprint in one dimension. It allows for reliable conclusions, about whether a parameter can be inferred from the experimental data. Three typical cases arise and can be detected from the profiles. A flat profile with a constant value indicates structural non-identifiability (c.f.~figure~\ref{ples}A). A profile that decreases but tails out to a plateau to one or both sides indicates practical non-identifiability (c.f.~figure~\ref{ples}B). A profile that tails out to zero on both sides quickly enough, i.e.~at least exponentially fast, indicates structural and practical identifiability (c.f.~figure~\ref{ples}C). Experimental design and model reduction strategies based on the profile likelihood allows one to resolve parameter non-identifiabilities iteratively, for an application see in \citep{Raue:2010fk}. 

Furthermore, the profile likelihood allows one to assess likelihood based confidence intervals \citep{Meeker:1995gd, Venzon:1988mk, Neale:1997zp}. A confidence interval $[\sigma_i^-,\,\sigma_i^+]$ to a confidence level $\alpha=0.95$ signifies that the true value of the parameter $\theta^*_i$ is expected to be inside the interval with 95\% probability. Using a threshold $\Delta_\alpha = \mathrm{Q}(\chi^2_{df},\,\alpha)$ which is the $\alpha$ quantile of the $\chi^{2}$--distribution with $df$ degrees of freedom \citep{Press:1990rw}, confidence intervals can be determined from the profiles by  $\{\theta_i \,\, | -2\cdot \log(PL(y|\theta_i)/L(y|\hat \theta)) < \Delta_\alpha\}$.

\subsection{Bayesian methods}
By applying Bayes' theorem the likelihood function (\ref{llhoodfun}) is extended by the prior probability density function (PDF) of the parameters $P(\theta)$ and normalised by a factor $c$ yielding the posterior PDF of the parameters
\begin{equation}
	P(\theta|y) = c\cdot L(y|\theta)\cdot P(\theta).
\end{equation}
In analogy to the MLE, the maximum a posteriori (MAP) estimate is defined by maximising $P(\theta|y)$. The only decisive difference to frequentist methodology is the choice of the prior $P(\theta)$. The direct computation of the normalisation factor $c$ is not feasible for a high dimensional parameter space. However, extensive Markov chain Monte Carlo (MCMC) sampling offers a way to evaluate $P(\theta|y)$ despite unknown $c$. This intriguing feature opens the prospect of considering the full high dimensional posterior PDF for statistical inference. Most prominently, the Metropolis-Hastings algorithm \citep{Metropolis:1953fk,Hastings:1970uq} defines a Markov process where transitions $\theta \rightarrow \theta'$ are generated using a proposal function $q(\theta|\theta')$ that eventually produces a series of samples of the posterior PDF $P(\theta|y)$. The transitions are accepted with probability 
\begin{equation}
	\alpha(\theta,\theta') = \min[1,\left(L(y|\theta')/L(y|\theta)\right)\cdot \left(q(\theta|\theta')/q(\theta'|\theta)\right)]. 
\end{equation}
For efficiency of the sampling the choice of the proposal function $q(\theta|\theta')$ is important. Often, proposals drawn from a multivariate normal distribution are convenient. One of the most simple implementations uses $q(\theta|\theta') \sim \mathrm{N}(0,s\cdot\mathbb{I})$ where $s\cdot\mathbb{I}$ is a scaled identity matrix (SIM). Too small a choice of $s$ in relation to the actual shape of the posterior PDF will cause the process to converge slowly and to produce correlated samples. Too large a choice of $s$ will lead to rejection of too many proposals $\theta'$ yielding a slow sampling. Assuming that the posterior PDF is a multivariate normal distribution the optimal acceptance rate of proposals is $\approx0.23$ \citep{Roberts:1997yq}. Nevertheless, in the light of complex and non-linear models with possibly limited amount and accuracy of experimental data, this assumption is problematic because the shape of the posterior PDF can be far from the PDF of a normal distribution. For a high dimensional parameter space computational efficiency also becomes an important issue. In these cases, the Markov chain may have to move along complex structures \citep[figure 8.2.2]{Box:1973kx}. To increase efficiency more sophisticated methods take into account the natural geometry of the posterior PDF, e.g.~the manifold Metropolis adjusted Langevin algorithm (MMALA) takes into account the local gradient and curvature information \citep{Girolami:2011uq}.

Before applying an MCMC sampling method, the prior PDF of the parameters needs to be specified. Given that empirical evidence exists about the distribution of a parameter, such as a previous measurements or estimation, $P(\theta)$ should incorporate this prior knowledge accordingly \citep{Efron:2005fk}. If no empirical evidence about the parameter value is available, the prior should be chosen as uninformative \citep{Box:1973kx}. This requires a flat metric in parameter space, i.e.~that does not artificially favour certain parameter values. Depending on the parametrisation of the model it can in practice be difficult to obtain a flat metric and hence to specify an uninformative prior \citep{Efron:2005fk}. 

The most crucial problem that MCMC sampling faces is to ensure that the samples obtained realistically represent the actual posterior PDF. One instance where the convergence of the Markov chain fails is if the posterior PDF is not proper \citep{Bayarri:2004uq}. Posterior PDF's are called proper if they are integrable \citep{Box:1973kx}. This means that it has to tail out to zero sufficiently fast over the appreciable range of the parameters such that its integral can be normalised to one. Non-identifiable parameters cause the posterior PDF to be non-proper \citep{Bayarri:2004uq}. This indicates that neither the prior assumptions nor the likelihood that represents the experimental data constrain the posterior PDF sufficiently. A practical consequence is that the Markov chain cannot converge and hence gives inaccurate results \citep{Gelfand:1999fk}. For convergence it is required that the Markov chain is positive recurrent which is not given in the case of non-identifiability \citep{Hobert:1996uq}. The user must ensure parameter identifiability before an MCMC technique can be used securely. If models contain many unknown parameters and the posterior PDF is not sufficiently constrained  the convergence of the Markov chain can be impractically slow even for a proper posterior PDF.

\section{Results}
\begin{figure} 
\begin{center}
\includegraphics[width=\textwidth]{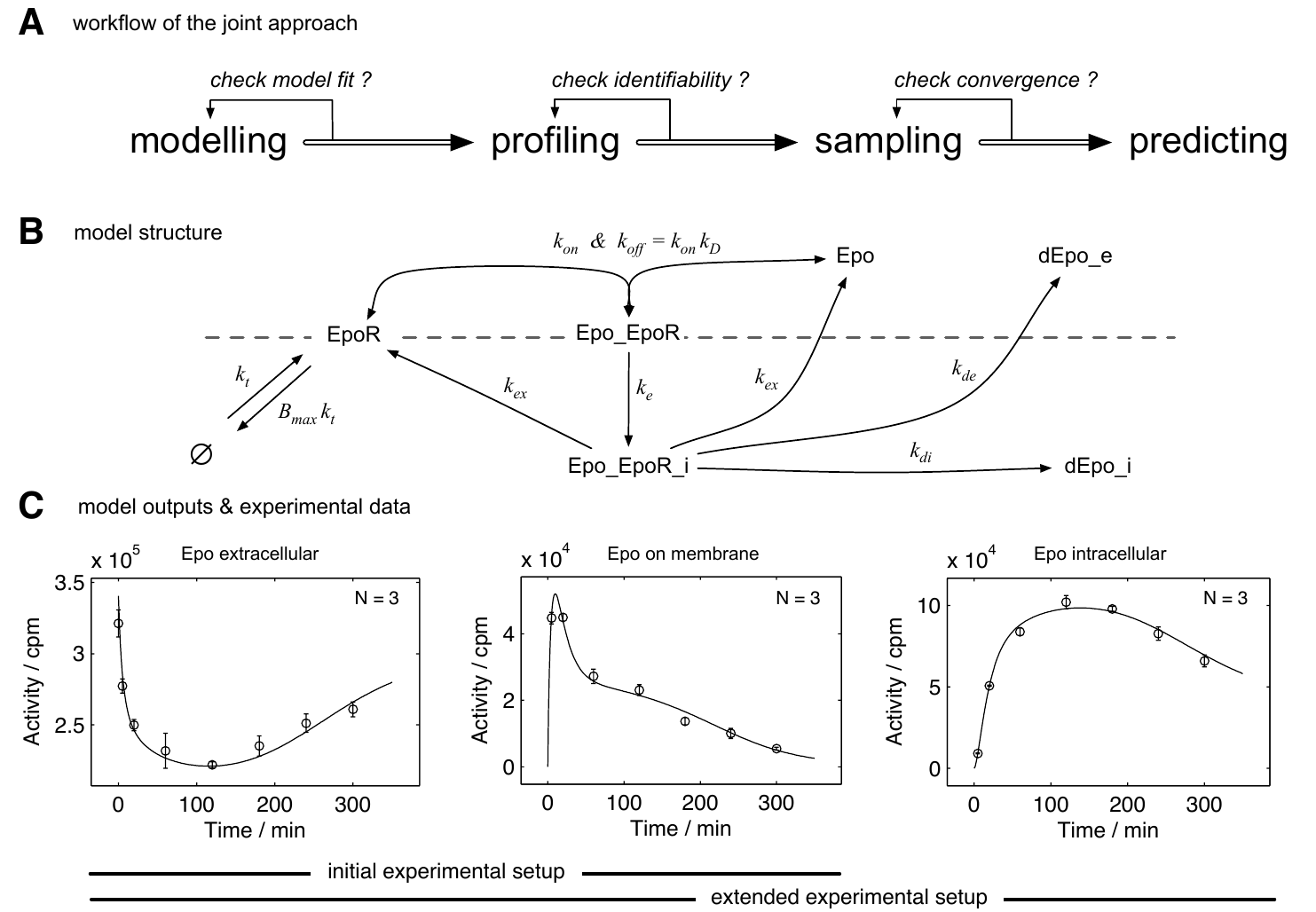}
\end{center}
\caption{(A) Workflow of the joint approach that combines profile likelihood and MCMC sampling methods: The first step is setting up an appropriate mathematical model (B) that contains the available prior knowledge and explains the experimental data sufficiently well (C). The profile likelihood approach allows one to detect and to resolve non-identifiability by experimental design. Therefore, it helps to ensure that the posterior PDF is proper and sufficiently well constrained. Subsequently, MCMC sampling can be applied securely. Finally, the uncertainty in model predictions can be assessed realistically using the obtained samples. (B) The figure shows the structure of the ODE model describing Epo and EpoR interactions. The dashed line indicates the cell membrane, the part above is the  outside, the part below is the inside of the cell. (C) The agreement of experimental data and model output for MAP parameter values for both experimental setups is displayed. The extended setup was derived by experimental design considerations, see in \citep{Raue:2010fk}.}
\label{workflow}
\end{figure}

For reliable inference in the presence of non-identifiability, we propose a joint approach that takes advantage of both profiling and MCMC sampling methods. The profile likelihood  is suitable to detect parameter non-identifiability \citep{Raue:2009ec}. Furthermore, it allows for experimental design that helps to resolve parameter non-identifiability \citep{Raue:2010fk}. This ensures that the posterior PDF is proper and well constrained. Subsequently, efficient MCMC sampling \citep{Girolami:2011uq} can be used reliably to generate samples of the posterior PDF. Finally, the uncertainty in model predictions can be assessed realistically. The workflow of this joint approach is displayed in figure~\ref{workflow}A. 

Both profiling and MCMC sampling methods separately are used for inference in various fields, e.g.~in particle physics \cite{Aprile:2011fk, Feroz:2011fk} or in climate research \cite{Moncho:2012fk, Smith:2009fk}. Here, we present results of using the joint approach that takes advantage of both methods for an application from cell biology \citep{Becker:2010hs}. An ordinary differential equation (ODE)  model was used to describe the concentration dynamics of six molecular compounds involved in the interplay of the hormone erythropoietin (Epo) and its corresponding membrane receptor (EpoR), see figure~\ref{workflow}B. Epo is an important factor in the differentiation of blood cells. The dynamics of the molecular compounds are formulated as
\begin{equation}
	\mathrm{d}\vec{x}(t,\theta)/\mathrm{d}t = \vec{f}(\vec{x}(t,\theta),\theta)
\end{equation}
and a mapping of the dynamics to experimentally accessible quantities by the model output function 
\begin{equation}
	\vec{y}(t,\theta) = \vec{g}(\vec{x}(t,\theta), \theta)
\end{equation}	
that enters the likelihood function (\ref{llhoodfun}) was used. The model comprises six ODEs and ten unknown parameters including one nuisance parameter, for details about the model equations see the supplementary notes. All parameters are positive by definition, hence the logarithmic space yields a flat metric \citep{Box:1973kx}. Since no empirical evidence about the values of the nine parameters of interest was available the prior was chosen uninformative. 

Radioactively labeled Epo facilitated the measurement of Epo concentration in the extracellular medium and of Epo bound to the cell membrane, see supplementary table~1. The MAP estimates of the model parameters were obtained by numerical optimisation, see supplementary table~2. The agreement of model outputs and experimental data for this initial experimental setup is shown in figure~\ref{workflow}C. In accordance with the experimental data the model describes the binding of Epo to the Epo receptor on the membrane, internalisation and recycling of the Epo-EpoR complex. As first step of the proposed joint approach the posterior PDF was screened for non-identifiability using the profiling method.

\subsection{Identifiability analysis using posterior profiles}
The profile likelihood approach for identifiability analysis \citep{Raue:2009ec} can be translated to a profile posterior approach. In analogy to equation (\ref{profilellh}) profiling can be applied to the unnormalised posterior PDF by defining the profile posterior
\begin{equation}
	PP(\theta_i|y) = \max_{\theta_{j\not=i}}[P(\theta|y)],
\end{equation}
c.f.~figure~\ref{ples}.  
Technical details on the implementation of the methods are given in the supplementary notes. For the initial experimental setup, the profiles of the posterior PDF reveal four structurally non-identifiable, two practically non-identifiable and three identifiable parameters. One of each case is displayed as red lines in figure~\ref{ple_vs_sampling}A, the remaining profiles are shown in supplementary figure~1. 

\begin{figure} 
\begin{center}
\includegraphics[width=\textwidth]{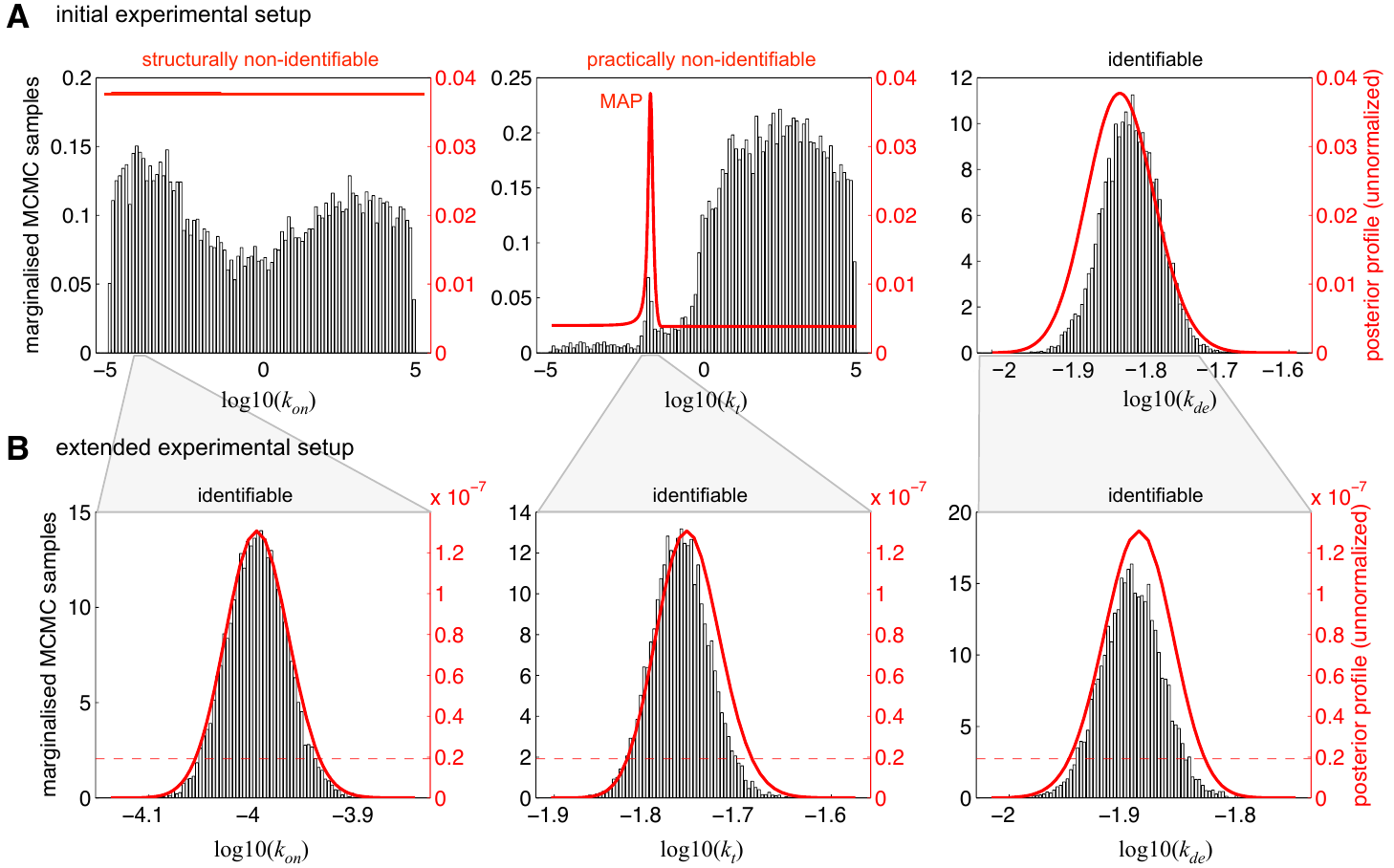}
\end{center}
\caption{Comparison of profiling and sampling results: (A) For the initial experimental setup the posterior profiles indicated by red lines reveal that parameter $k_{on}$ is structurally non-identifiable, parameter $k_t$ is practically non-identifiable and parameter $k_{de}$ is identifiable. The histograms display the marginalised MCMC samples obtained by the MMALA algorithm. For the identifiable parameter $k_{de}$ both results of profiling and sampling agree quite well. Also for the structurally non-identifiable parameter $k_{on}$ the agreement is acceptable. For the practically non-identifiable parameter $k_t$ the results are substantially different. The profile shows that the MAP point is located at $\log_{10}(k_t) \approx -1.8$. However, the lion's share of the marginalised MCMC samples propose $\log_{10}(k_t)$ to be $>0$. (B) Taking into account more experimental data the posterior profiles for the extended experimental setup indicate that all parameters are now identifiable. The results of MCMC sampling and profiling are in good agreement. Interestingly, the MCMC samples for parameter $k_t$ for the extended setup are localised close to the MAP point of the initial setup, note the different scales on the x-axis for (A) and (B). The dashed red lines indicates the threshold $\Delta_\alpha$ that can be used to assess likelihood based confidence intervals.}
\label{ple_vs_sampling}
\end{figure}

\subsection{Markov chain Monte Carlo sampling}
The results of the posterior profiling approach indicates that the posterior PDF for the initial experimental setup is not proper, i.e.~the posterior PDF cannot be normalised to one and is therefore not a valid PDF. In this situation an MCMC sampling cannot converge and hence gives inaccurate results \citep{Gelfand:1999fk}. In order to allow for a comparison between profiling and MCMC sampling, the prior PDF was restricted artificially to a uniform distribution with support from $10^{-5}$ to $10^{+5}$. The resulting posterior PDF is now proper and the Markov chain can in principle converge. It is important to note that the results of MCMC sampling can potentially be biased due to the artificial specification of this prior PDF. The new posterior PDF, though proper, still exhibits a complicated structure. The probability mass is distributed widely in the ten dimensional parameter space, see the posterior profiles in figure~\ref{ple_vs_sampling}A. In this situation, the SIM sampling algorithm, that uses a scale identity matrix for generating proposals, was too inefficient and did not yield reasonable results within an acceptable time, see supplementary figures~1--3. To improve efficiency the MMALA algorithm that takes into account the local geometry of $P(\theta|y)$ was used \citep{Girolami:2011uq}. Figure \ref{ple_vs_sampling}A shows the histograms of marginalised MCMC samples. The remaining results are displayed in supplementary figures~4--6. For the identifiable parameter $k_{de}$ and for the structural non-identifiable parameter $k_{on}$ the results of the MCMC sampling are in good agreement with the results of the profiling. For the practically non-identifiable parameter $k_t$ the results are substantially different. The profile shows that the MAP point is located at $\log_{10}(k_t) \approx -1.8$. However, the lion's share of the marginalised MCMC samples propose $\log_{10}(k_t)$ to be $>0$. In this region the posterior profile reveals a plateau.

\subsection{Taking into account additional experimental data.}
Both the results of posterior profiling and MCMC sampling indicate substantial uncertainty in the parameter estimates for the initial experimental setup. Based on the results of the profiling approach additional experiments were suggested, see in \citep{Raue:2010fk} for details. The target of the experimental design was to resolve non-identifiabilities. The additional experimental data were included in the estimation procedure, yielding an extended experimental setup, see figure~\ref{workflow}C. Figure~\ref{ple_vs_sampling}B shows the recomputed posterior profiles. They indicate, by tailing out to zero, that the identifiability problems were resolved for all parameters. The remaining profiles are shown in supplementary figure~7. 

As a consequence the posterior PDF is now proper without artificial assumptions on the prior PDF. In this situation, the SIM sampling algorithm was still too inefficient and did not yield reasonable results within an acceptable time, see supplementary figures~7--9. To improve efficiency the MMALA algorithm that takes into account the local geometry of $P(\theta|y)$ was used again \citep{Girolami:2011uq}. Figure \ref{ple_vs_sampling}B shows the histograms of marginalised MCMC samples. The results for the remaining parameters are shown in supplementary figures~10--12. For all parameters the results of the MCMC sampling and posterior profiling are now in good agreement. The MCMC samples for parameter $k_t$ that showed substantial difference between profiles and MCMC samples for the initial setup are now in good agreement as well. Interestingly, the MCMC samples for parameter $k_t$ for the extended setup are localised close to the MAP point of the initial setup. This suggests that the large probability mass of MCMC samples for values of $\log_{10}(k_t)>0$ obtained for the initial setup was misleading. 

\begin{figure} 
\begin{center}
\includegraphics[width=\textwidth]{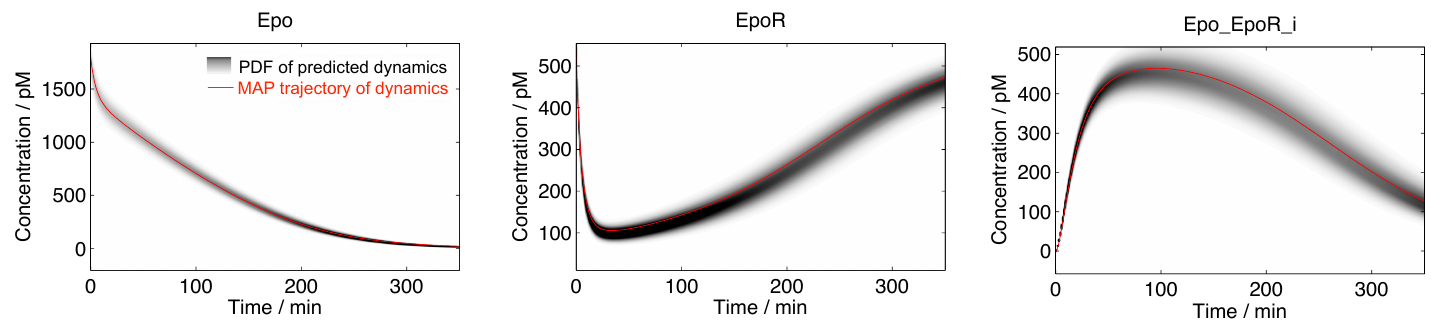}
\end{center}
\caption{Propagation of parameter uncertainties to model predictions: The posterior samples that were acquired by MCMC sampling for the extended setup, see figure~\ref{ple_vs_sampling}B, can now be used to assess prediction uncertainties accurately. The figure shows the predicted dynamics of three molecular components that could not be measured directly. The red line shows the prediction that corresponds to the MAP estimate of the parameters. The grey scale displays the posterior PDF of the prediction.}
\label{predictions}
\end{figure}

MCMC sampling results are now reliable and allow one to consider the full high dimensional posterior PDF for inference, see e.g.~the non-linear correlation between parameter $k_{di}$ and $k_{de}$ shown in supplementary figure~13. Using the generated MCMC samples, the parameter uncertainties contained in the posterior PDF can be propagated accurately to the prediction of the dynamics of molecular compounds that are not accessible by experiments directly, see figure~\ref{predictions}.

\section{Summary}
We introduced a joint approach that combines frequentist profiling and efficient Bayesian MCMC sampling methods. The proposed approach uses the analysis of posterior profiles that helps to determine the identifiability of the model parameters. Non-identifiability results in a posterior distribution that is not proper, i.e.~that cannot be normalised to one. However, a proper posterior distribution is required for convergence of the MCMC sampling. The results of the posterior profiling can be used for experimental design that helps resolve non-identifiabilities iteratively. Consequently, this ensures that the posterior distribution is proper. Having ensured identifiability of the parameters, the MCMC sampling results are reliable and can be used to propagate the uncertainty of parameter estimates to model predictions. Using an application from cell biology we showed that this approach enables one to obtain accurate results.

We compared the results of posterior profiling and marginalising over the MCMC samples for two stages of experimental setup: an initial setup that contains non-identifiabilities and an extended setup where all parameters are identifiable. A uniform prior distribution ensured that the posterior is proper despite non-iden\-ti\-fi\-a\-bi\-li\-ty for the initial setup. For the extended setup the results of posterior profiling and marginalising over the MCMC samples are in good agreement. For the initial setup substantial difference between profiles and MCMC samples are observed. Interestingly, the profiles for the initial setup reflect the posterior distribution for the extended setup much better than the MCMC samples of the initial setup. This indicates that MCMC sampling results in the presence of non-identifiability can be inaccurate despite a proper posterior distribution. 

\begin{acknowledgements}
This work was supported by the German Federal Ministry of Education and Research [Virtual Liver (Grant No. 0315766), LungSys (Grant No. 0315415E), and FRISYS (Grant No. 0313921)], the European Union [CancerSys (Grant No. EU-FP7 HEALTH-F4-2008-223188)], the Initiative and Networking Fund of the Helmholtz Association within the Helmholtz Alliance on Systems Biology (CoReNe HMGU), and the Excellence Initiative of the German Federal and State Governments (EXC 294)
\end{acknowledgements}

\bibliographystyle{unsrtnat}
\bibliography{template-reduces.bbl}

\label{lastpage}
\end{document}


\title[Supplement: Joining Bayesian and Frequentist Methods]{Supplementary Notes and Figures for Joining Forces of Bayesian and Frequentist Methodology: A Study for Inference in the Presence of Non-Identifiability}

\author[A. Raue and others]{Andreas Raue$^{1}$, Clemens Kreutz$^{1}$, \\Fabian Joachim Theis$^2$ and Jens Timmer$^{1,3}$}

\affiliation{$^1$ Institute for Physics, University of Freiburg, Germany\\
$^2$ Helmholtz Zentrum Munich and Departement of Mathematics, Technical University of Munich, Germany\\
$^3$ BIOSS Centre for Biological Signalling Studies and Freiburg Institute for Advanced Studies (FRIAS), Freiburg, Germany; Department of Clinical and Experimental Medicine, Linköping University, Sweden}

\label{firstpage}

\maketitle

\section{Model description and experimental setup.} 
The mathematical model presented in \citep{Becker:2010hs} describes the concentration dynamics of molecular compounds $\vec{x}$ involved in the interplay of the hormone erythropoietin (Epo) and its corresponding membrane receptor (EpoR). Epo is an important factor in the differentiation of blood cells. The dynamics of the molecular compounds are formulated as ordinary differential equations (ODE) $\mathrm{d}\vec{x}(t,\theta)/\mathrm{d}t = \vec{f}(\vec{x}(t,\theta),\theta)$ and a mapping of the compounds to experimentally accessible quantities by a function $\vec{y}(t,\theta) = \vec{g}(\vec{x}(t,\theta), \theta)$ is used. The model comprises six dynamical variables and ten unknown parameters including one nuisance parameter. The dynamics are given by the ODE system
\begin{eqnarray}
	\mathrm{d}/\mathrm{d}t\,{x_1} & = &  - {k_{on}} \cdot x_1 \cdot  x_2 + {k_{on}} \cdot{k_D} \cdot x_3 + {k_{ex}} \cdot x_4 \nonumber \\
	\mathrm{d}/\mathrm{d}t\,{x_2} & = &  - {k_{on}} \cdot x_1 \cdot  x_2 + {k_{on}} \cdot{k_D} \cdot x_3 + {k_t} \cdot{B_{max}} - {k_t} \cdot x_2 + {k_{ex}} \cdot x_4 \nonumber \\
	\mathrm{d}/\mathrm{d}t\,{x_3} & = &  + {k_{on}} \cdot x_1 \cdot  x_2 - {k_{on}} \cdot{k_D} \cdot x_3 - {k_e} \cdot x_3 \nonumber \\
	\mathrm{d}/\mathrm{d}t\,{x_4} & = &  + {k_e} \cdot x_3 - {k_{ex}} \cdot x_4 - {k_{di}} \cdot x_4 - {k_{de}} \cdot x_4 \nonumber \\
	\mathrm{d}/\mathrm{d}t\,{x_5} & = &  + {k_{di}} \cdot x_4 \nonumber \\
	\mathrm{d}/\mathrm{d}t\,{x_6} & = &  + {k_{de}} \cdot x_4 \nonumber	
\end{eqnarray}
that describes the time evolution of the concentration of molecular compounds. The modelled compounds are: Epo in the extracellular medium ($x_1$, Epo), Epo receptor on the cell membrane ($x_2$, EpoR), occupied receptor complex on the cell membrane ($x_3$, Epo\_EpoR), internalized receptor complex ($x_4$, Epo\_EpoR\_i), degraded Epo inside the cell ($x_5$, dEpo\_i) and degraded Epo in the extracellular medium ($x_6$, dEpo\_e). It is assumed that two compounds have non zero initial value defined as unknown parameters: $x_1(t=0) = Epo_\mathit{0}$ and $x_2(t=0) = B_{max}$. The molecular interaction that are considered are: ligand-independent EpoR endocytosis ($k_t$), association of Epo and EpoR ($k_{on}$), dissociation of Epo and EpoR ($k_{off} = k_{on}\cdot k_D$) with dissociation constant for Epo\_EpoR ($k_D$), ligand-induced EpoR endocytosis ($k_e$), recycling of Epo and EpoR ($k_{ex}$), degradation of ligand-EpoR complex that remaining intracellular ($k_{di}$) and is secreted extracellular ($k_{de}$). The experimentally accessible quantities are given by
\begin{eqnarray}
	y_1 & = & scale \cdot (x_1 + x_6)  \nonumber \\
	y_2 & = & scale \cdot x_3  \nonumber \\
	y_3 & = & scale \cdot (x_4 + x_5)  \nonumber
\end{eqnarray}
and describe measurements of radioactively labeled Epo in different compartments: in the extracellular medium ($y_1$, Epo\_ext), on the cell membrane ($y_2$, Epo\_mem) and inside the cell ($y_3$, Epo\_int), see table~\ref{data} for data. The mathematical model and the experimental data is available online from the original publication \citep{Becker:2010hs} in standard formats. 

For the analysis two experimental setups are distinguished. The initial experimental setup includes only measurements of $y_1$ and $y_2$ and results in non-identifiability, see posterior profiles in figure~\ref{results1_naiv}. Based on the results of the profiling approach additional experiments were suggested, see in \citep{Raue:2010fk} for details. The target of the experimental design was to resolve non-identifiabilities. Additional measurements of $y_3$ and independent direct measurements of $B_{max}$, $k_D$ and $Epo_0$ were included in the estimation procedure. The posterior profiles computed for the extended setup confirm that the non-identifiabilities were resolved, see in figure~\ref{results2_naiv}. The MAP parameter values for the extended setup are given in table~\ref{map_pars}. 

For MCMC sampling both initial and extended setup have been analysed by the SIM and MMALA algorithms. The SIM algorithm uses a scaled identity matrix for generating proposals. To improve efficiency the MMALA algorithm that takes into account the local geometry of the posterior PDF \citep{Girolami:2011uq}. 1) For the initial setup the results of the SIM algorithm are displayed in figures~\ref{results1_naiv}--\ref{chains1_naiv}. They show that the SIM algorithm produced correlated results and did not converge yet. 2) For the initial setup the results of the MMALA algorithm are displayed in figure~\ref{results1_suppl}--\ref{chains1}. The MMALA algorithm converged. 3) For the extended setup the results of the SIM algorithm are displayed in figures~\ref{results2_naiv}--\ref{chains2_naiv}. They show that the SIM algorithm still produced correlated results and did not converge yet. 4) For the extended setup the results of the MMALA algorithm are displayed in figure~\ref{results2_suppl}--\ref{chains2}. The MMALA algorithm converged.

\section{Implementation of the numerical methods.} 
\begin{figure}[b]
\begin{center}
\includegraphics[width=\textwidth]{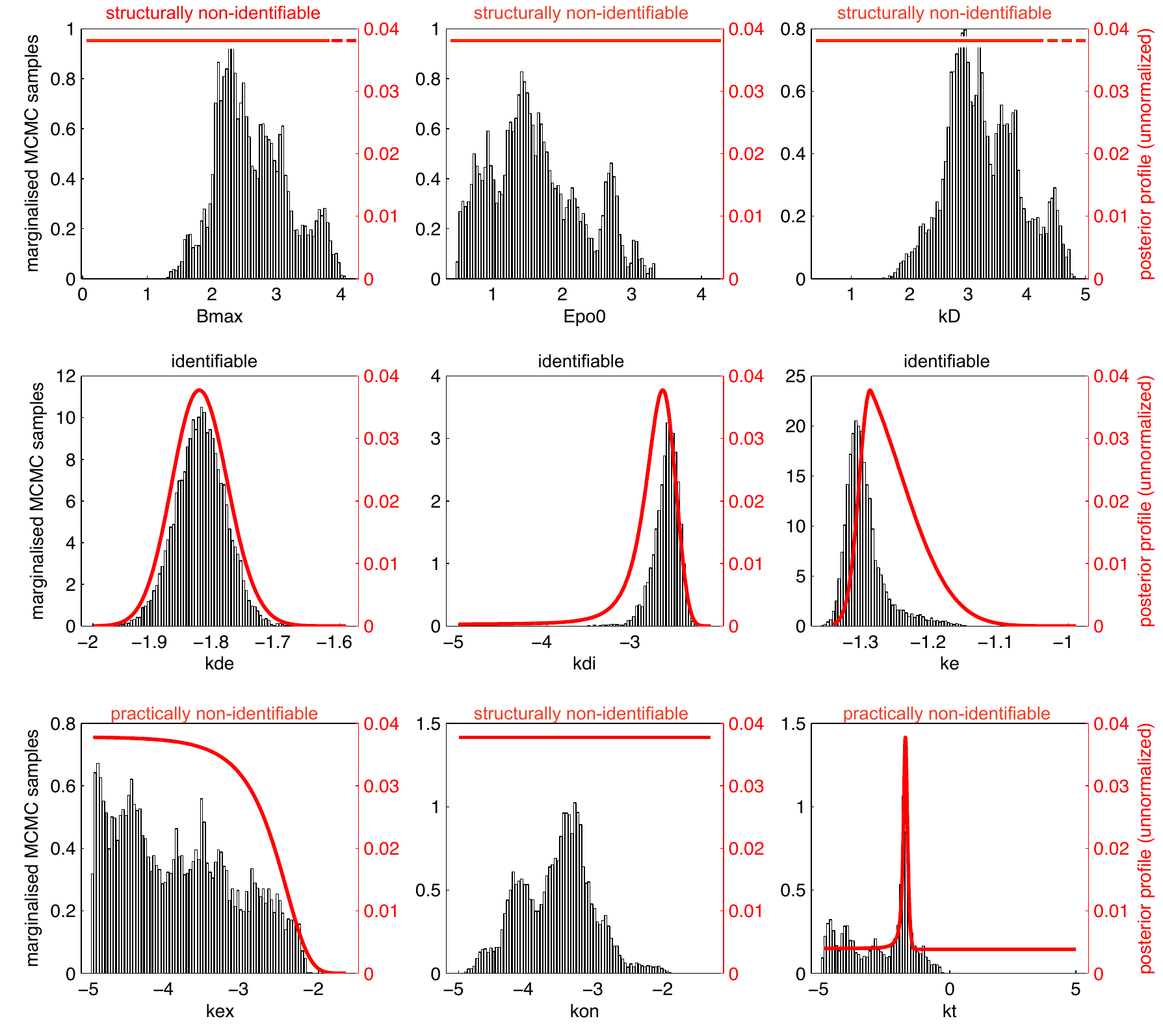}
\end{center}
\caption{Results of the posterior profiling and MCMC sampling using the SIM algorithm for the initial setup: In contrast to the results obtained by the simplified MMALA algorithm shown in figure~\ref{results1_suppl}, the SIM algorithm converges too slowly for parameters affected by non-identifiability. The sampling results are of bad quality. $10^6$ MCMC samples were generated and a thinning of 1/100 was used. The auto-correlation function of the samples still shows high correlation, see figure~\ref{acf1_naiv}. This indicates that the process did not converge yet. The Markov chains for each parameter are displayed in figure~\ref{chains1_naiv}.}
\label{results1_naiv}
\end{figure}

\begin{figure}[b]
\begin{center}
\includegraphics[width=\textwidth]{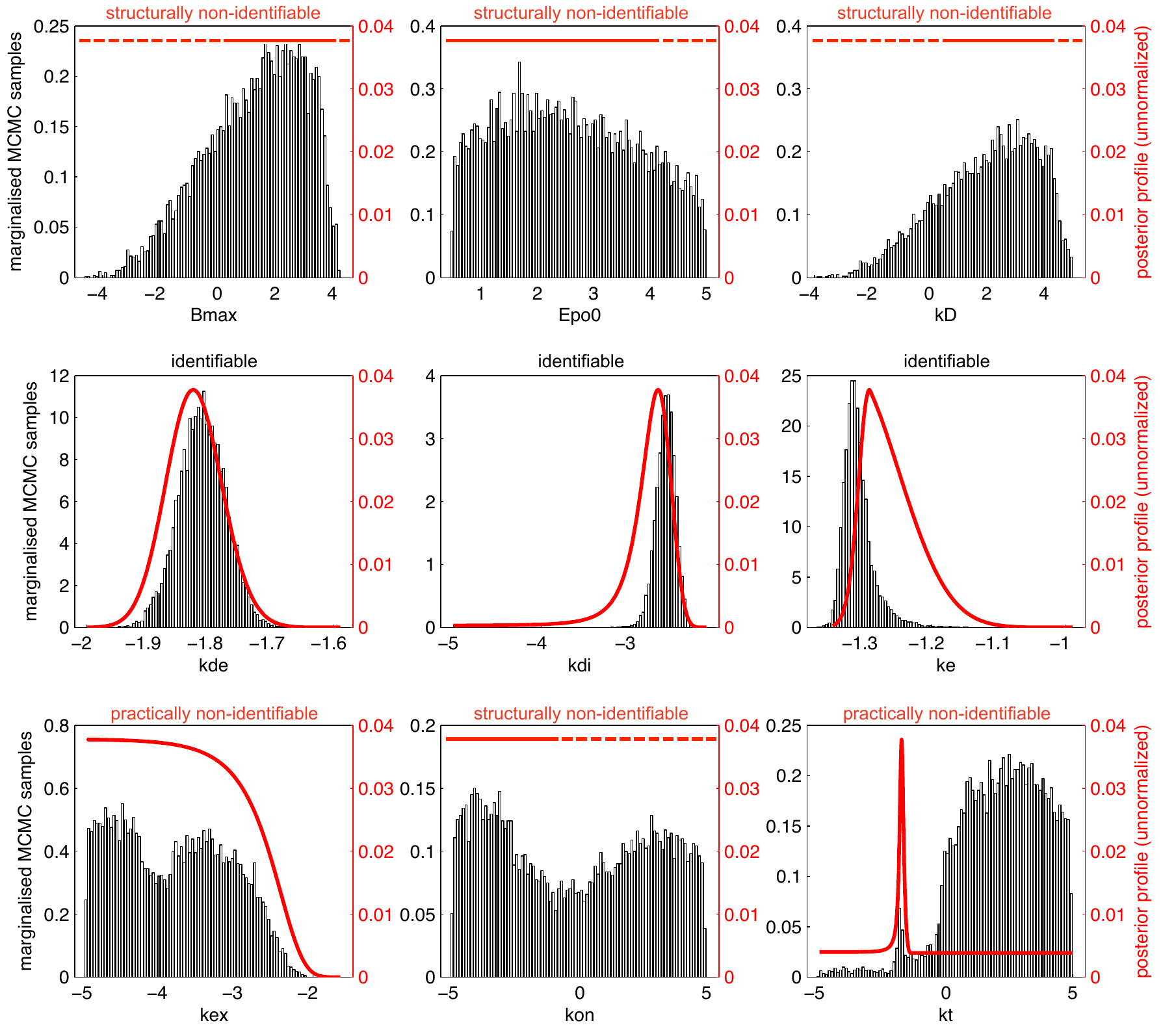}
\end{center}
\caption{Results of the posterior profiling and MCMC sampling using the MMALA algorithm for the initial setup: For most parameters the results of profiling and MCMC sampling are similar though not identical. For parameter $k_t$ substantial difference are observed. $10^6$ MCMC samples were generated. A thinning of 1/100 was. The independence of samples was checked by their auto-correlation function, see figure~\ref{acf1}. This indicates that the process did converge. The Markov chains for each parameter are displayed in figure~\ref{chains1}.}
\label{results1_suppl}
\end{figure}

\begin{figure}[b]
\begin{center}
\includegraphics[width=\textwidth]{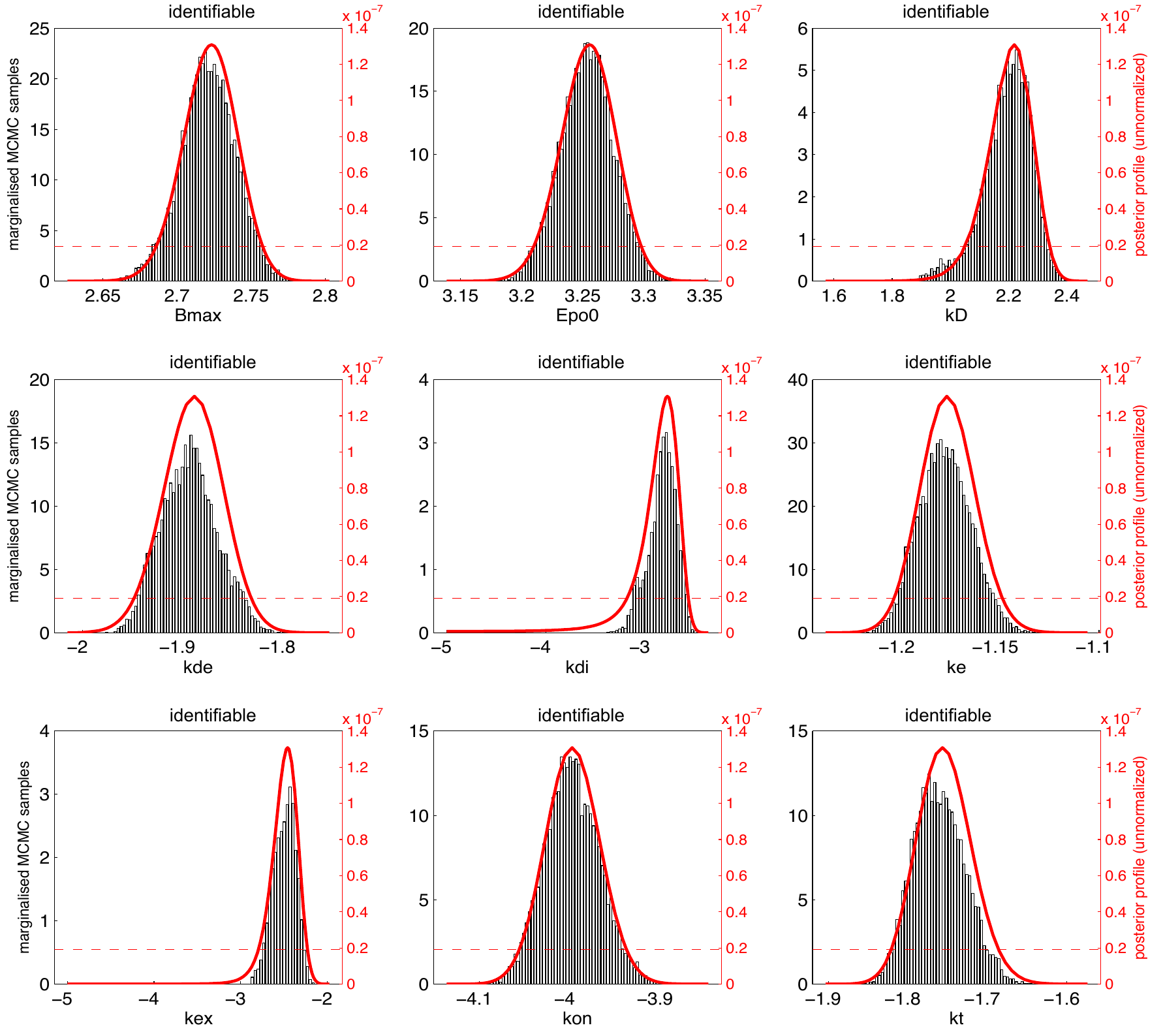}
\end{center}
\caption{Results of the posterior profiling and MCMC sampling using the SIM algorithm for the extended setup: In contrast to the results obtained by the simplified MMALA algorithm shown in figure~\ref{results2_suppl}, the SIM algorithm converges too slowly. Nevertheless, the agreement of profiles and MCMC samples is already acceptable. $10^6$ MCMC samples were generated and a thinning of 1/100 was used. The auto-correlation function of the samples still shows high correlation, see figure~\ref{acf2_naiv}. The Markov chains for each parameter are displayed in figure~\ref{chains2_naiv} and indicate that the process did not converge yet. The dashed red lines indicates the threshold $\Delta_\alpha$ that can be used to assess likelihood based confidence intervals.}
\label{results2_naiv}
\end{figure}

\begin{figure}[b]
\begin{center}
\includegraphics[width=\textwidth]{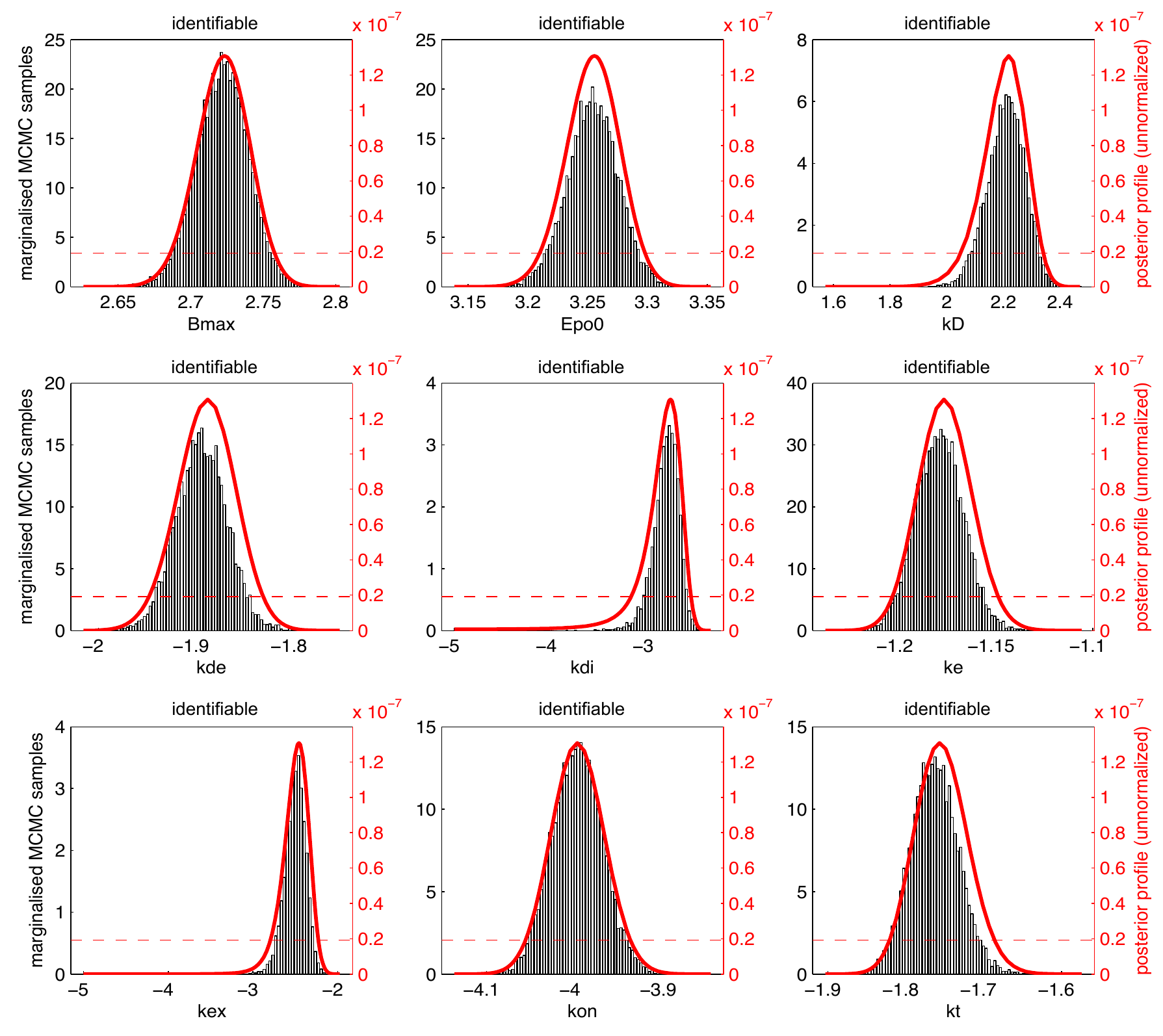}
\end{center}
\caption{Results of the posterior profiling and MCMC sampling using the MMALA algorithm for the extended setup: the results of profiling and MCMC sampling are in good agreement. $10^5$ MCMC samples were generated. A thinning of 1/10 was used. The independence of samples was checked by their auto-correlation function, see figure~\ref{acf2}. This indicates that the process did converge. The Markov chains for each parameter are displayed in figure~\ref{chains2}. The dashed red lines indicates the threshold $\Delta_\alpha$ that can be used to assess likelihood based confidence intervals.}
\label{results2_suppl}
\end{figure}

The ODE system was solved by the CVODES algorithm \citep{Hindmarsh:2005fb}. For numerical optimisation the trust-region method LSQNONLIN from MATLAB was used yielding the MAP estimates \citep{Coleman:1996fk}. For efficiency it takes into account local gradient and curvature information. For the calculation of the posterior profiles as shown in figures \ref{results1_naiv}, \ref{results1_suppl}, \ref{results2_naiv} and \ref{results2_suppl} an algorithm described in \citep{Raue:2009ec} can be used. 

Both MMALA and LSQNONLIN rely on the accuracy of first order sensitivities $\mathrm{d}\vec{y}/\mathrm{d}\theta$. For ODE systems finite difference should not be used to calculate sensitivities \citep{Conn:2009kx}. Therefore the sensitivity equations \citep{Leis:1988dl} are solved simultaneously by the ODE solver CVODES. For the implementation of the MMALA algorithm, the simplified version was used that is computationally more efficient \citep{Girolami:2011uq}. 

\begin{figure}[b]
\begin{center}
\includegraphics[width=\textwidth]{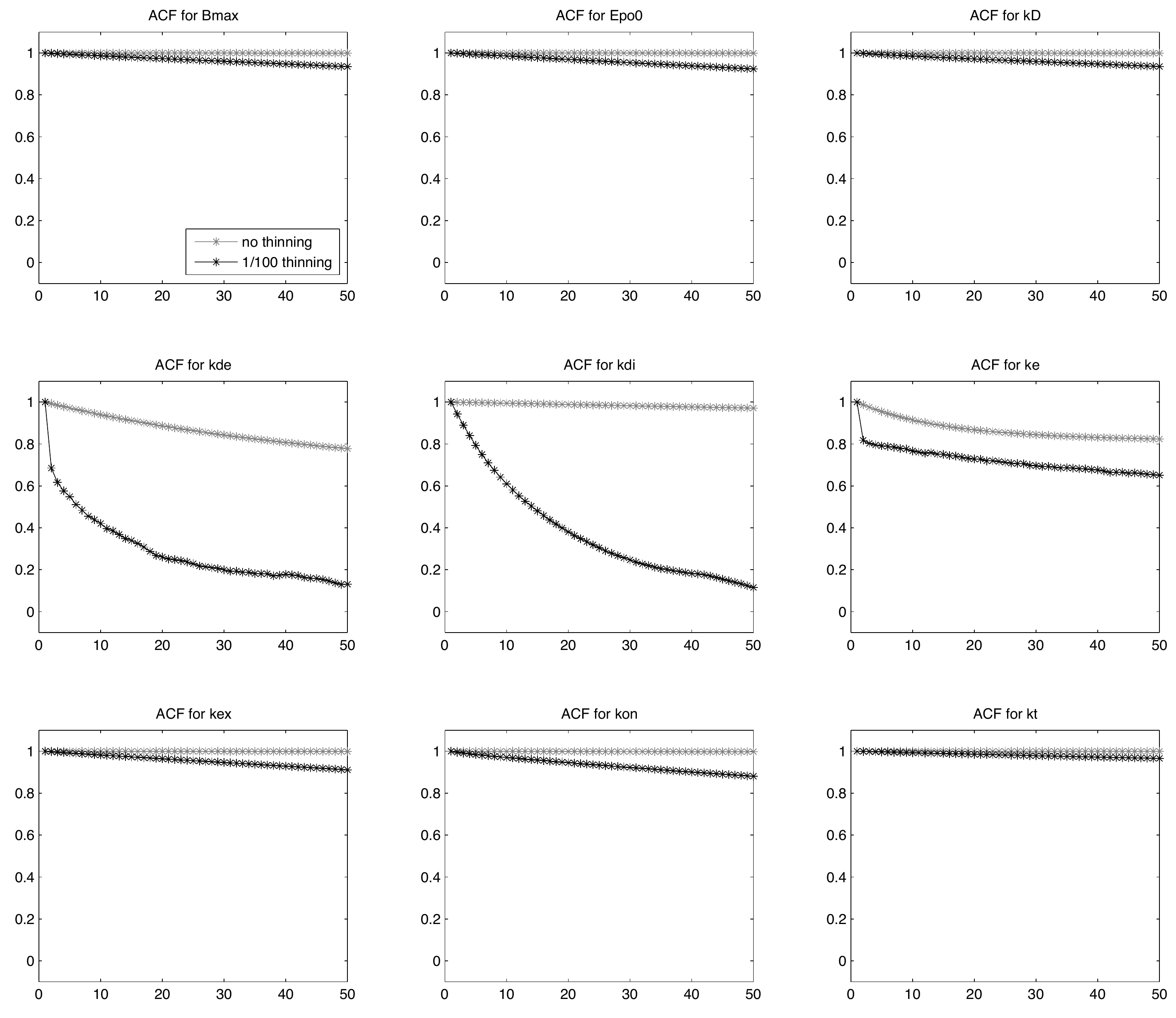}
\end{center}
\caption{ACF of the Markov chain using the SIM algorithm for the initial setup: The figure displays the auto-correlation function (ACF) of the generated MCMC samples for each parameter along the horizontal axis. A thinning of 1/100 was used to increase independence of the samples. A rapidly decaying ACF indicates independence of the samples and convergence of the Markov chain. The sample are still highly correlated.}
\label{acf1_naiv}
\end{figure}

\begin{figure}[b]
\begin{center}
\includegraphics[width=\textwidth]{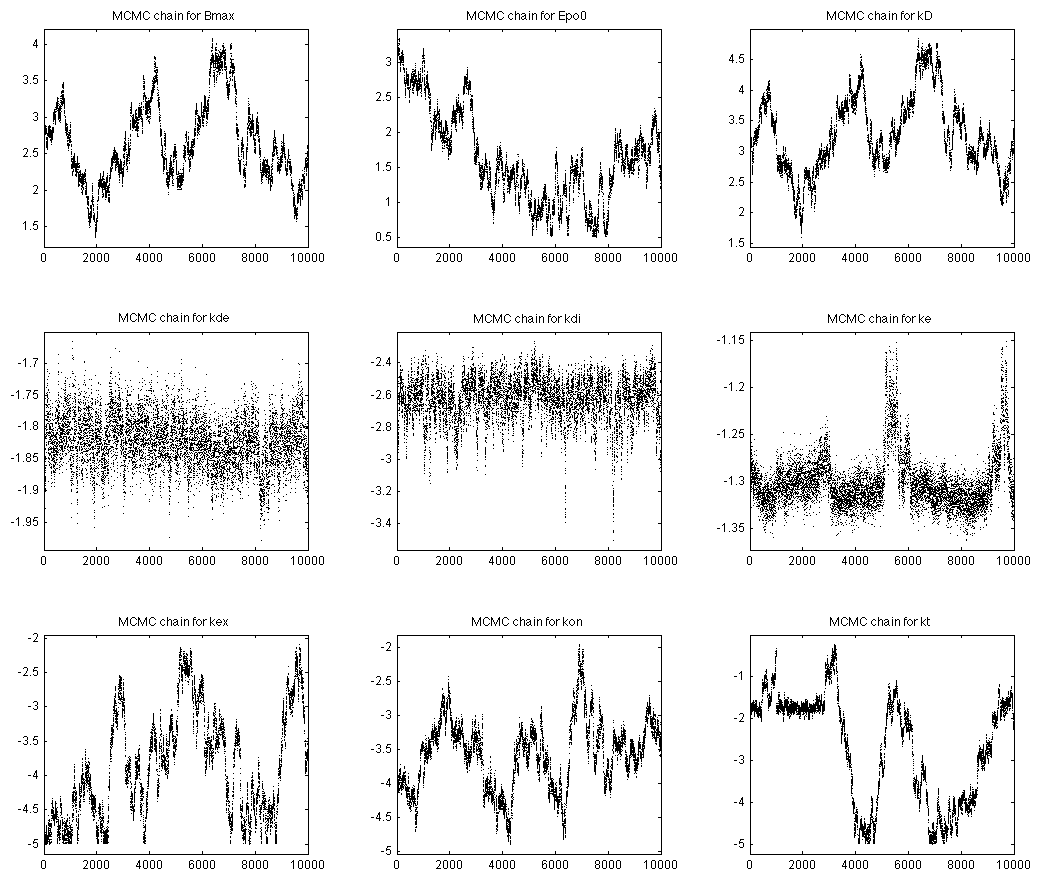}
\end{center}
\caption{Markov chains for the initial setup using the SIM algorithm: The figure displays the generated samples for each parameter after thinning along the horizontal axis. For parameters affected by non-identifiability, i.e. parameters in the upper and lower row, the Markov process did not converge after $10^4$ samples were generated. In contrast to the results obtained by the simplified MMALA algorithm the SIM algorithm converges too slowly.}
\label{chains1_naiv}
\end{figure}

\begin{figure}[b]
\begin{center}
\includegraphics[width=\textwidth]{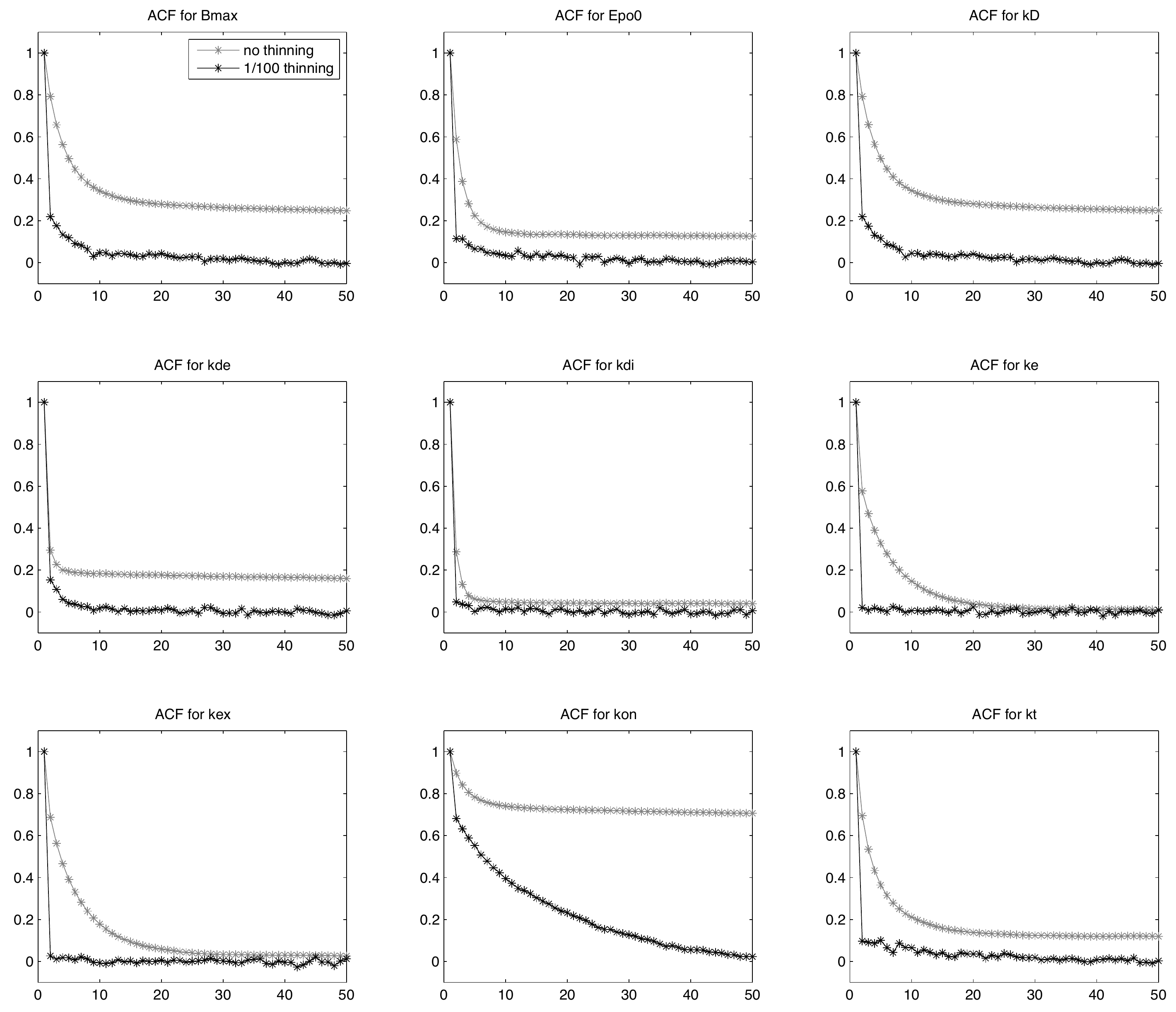}
\end{center}
\caption{ACF of the Markov chain using the MMALA algorithm for the initial setup: The figure displays the auto-correlation function (ACF) of the generated MCMC samples for each parameter along the horizontal axis. A thinning of 1/100 was used to increase independence of the samples. A rapidly decaying ACF indicates independence of the samples and convergence of the Markov chain. For parameter $k_{on}$ some dependency remained. However, the ACF drops to a value close to zero within a lag of 50 which is much smaller than the number of MCMC samples generated.}
\label{acf1}
\end{figure}

\begin{figure}[b]
\begin{center}
\includegraphics[width=\textwidth]{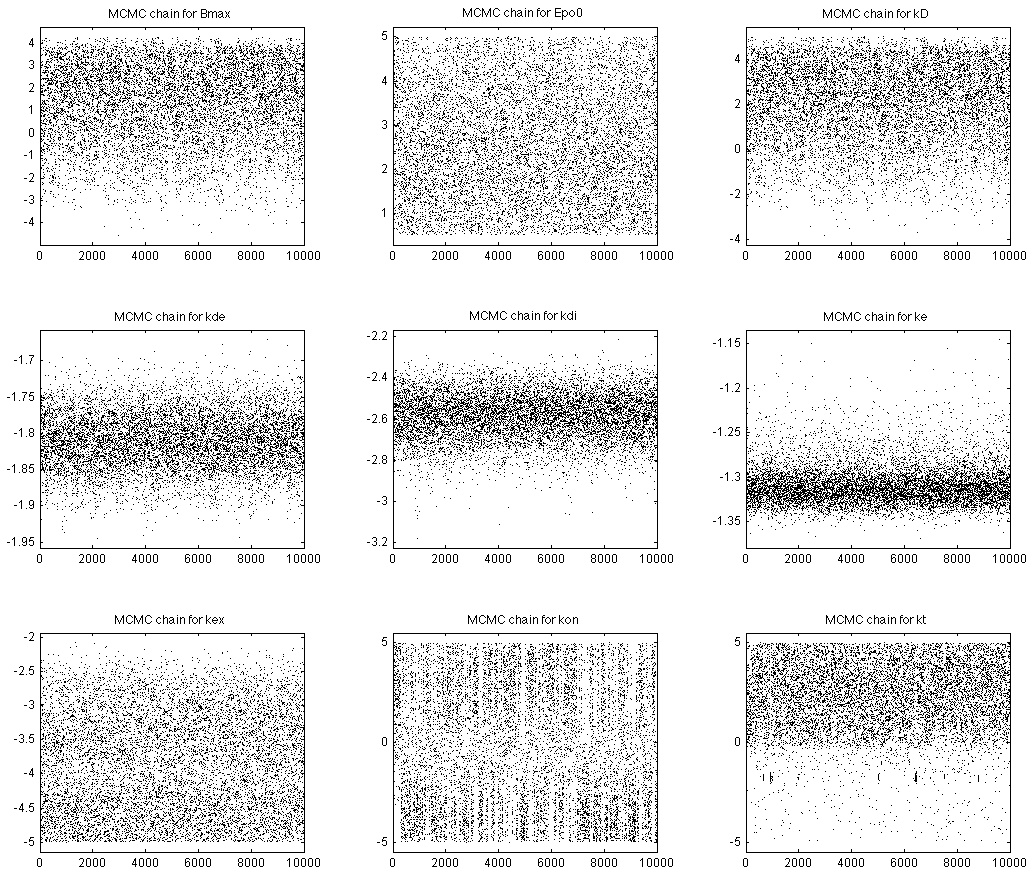}
\end{center}
\caption{Markov chains using the MMALA algorithm for the initial setup: The figure displays the generated samples for each parameter after thinning along the horizontal axis. }
\label{chains1}
\end{figure}

\begin{figure}[b]
\begin{center}
\includegraphics[width=\textwidth]{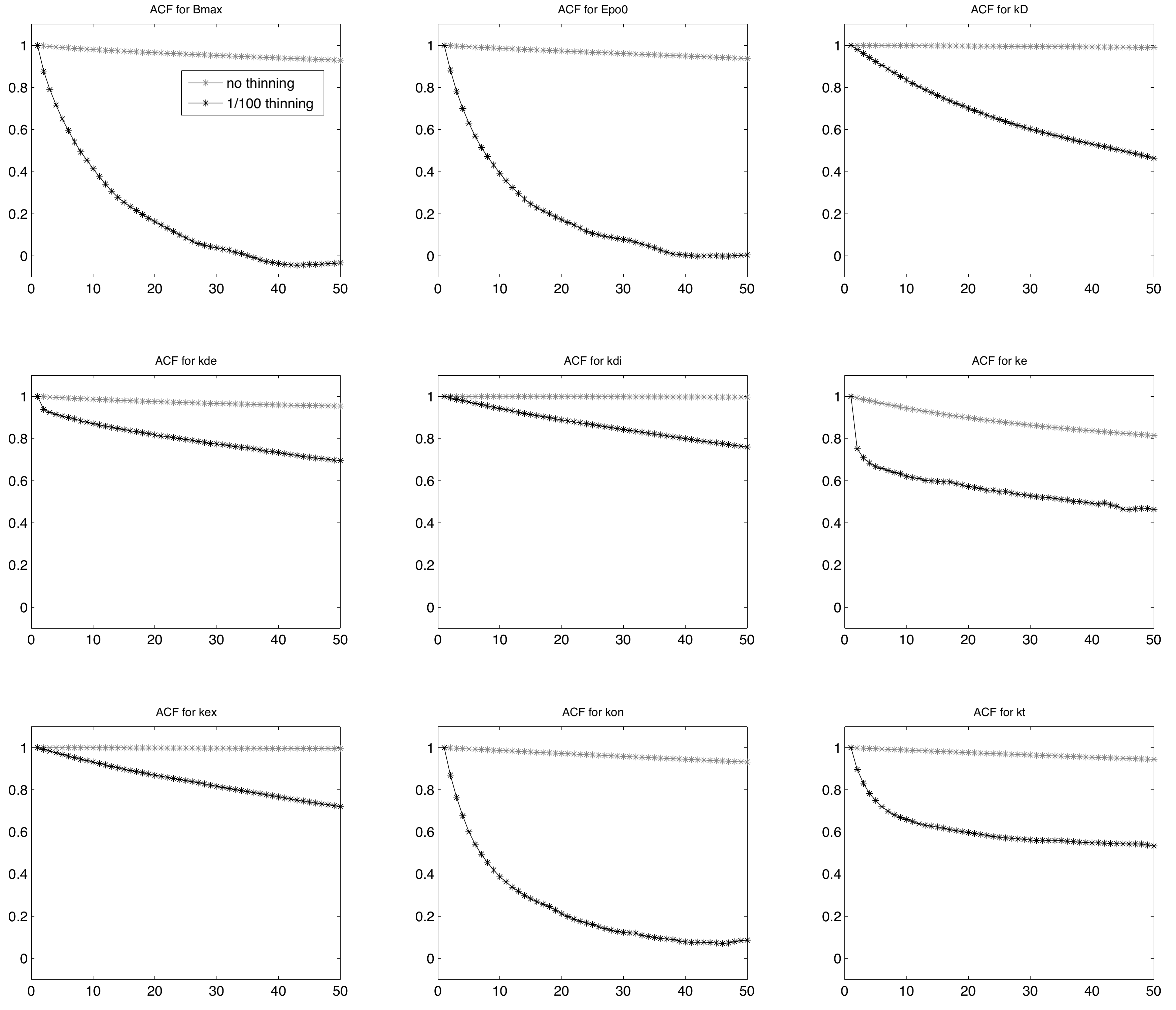}
\end{center}
\caption{ACF of the Markov chain using the SIM algorithm for the extended setup: The figure displays the auto-correlation function (ACF) of the generated MCMC samples for each parameter along the horizontal axis. A thinning of 1/100 was used to increase independence of the samples. A rapidly decaying ACF indicates independence of the samples and convergence of the Markov chain. However the sample are still highly correlated.}
\label{acf2_naiv}
\end{figure}

\begin{figure}[b]
\begin{center}
\includegraphics[width=\textwidth]{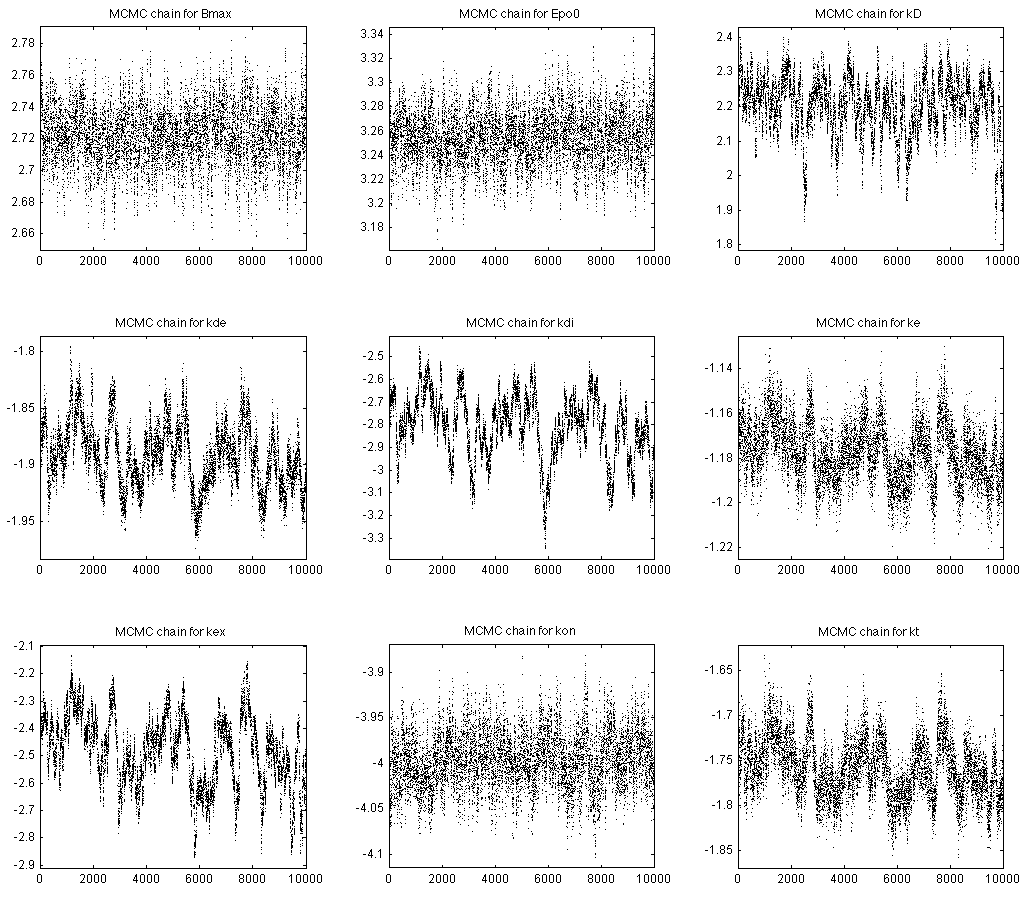}
\end{center}
\caption{Markov chains for the extended setup using the SIM algorithm: The figure displays the generated samples for each parameter after thinning along the horizontal axis. For some parameters the Markov process did not converge after $10^4$ samples were generated. In contrast to the results obtained by the simplified MMALA algorithm the SIM algorithm converges too slowly.}
\label{chains2_naiv}
\end{figure}

\begin{figure}[b]
\begin{center}
\includegraphics[width=\textwidth]{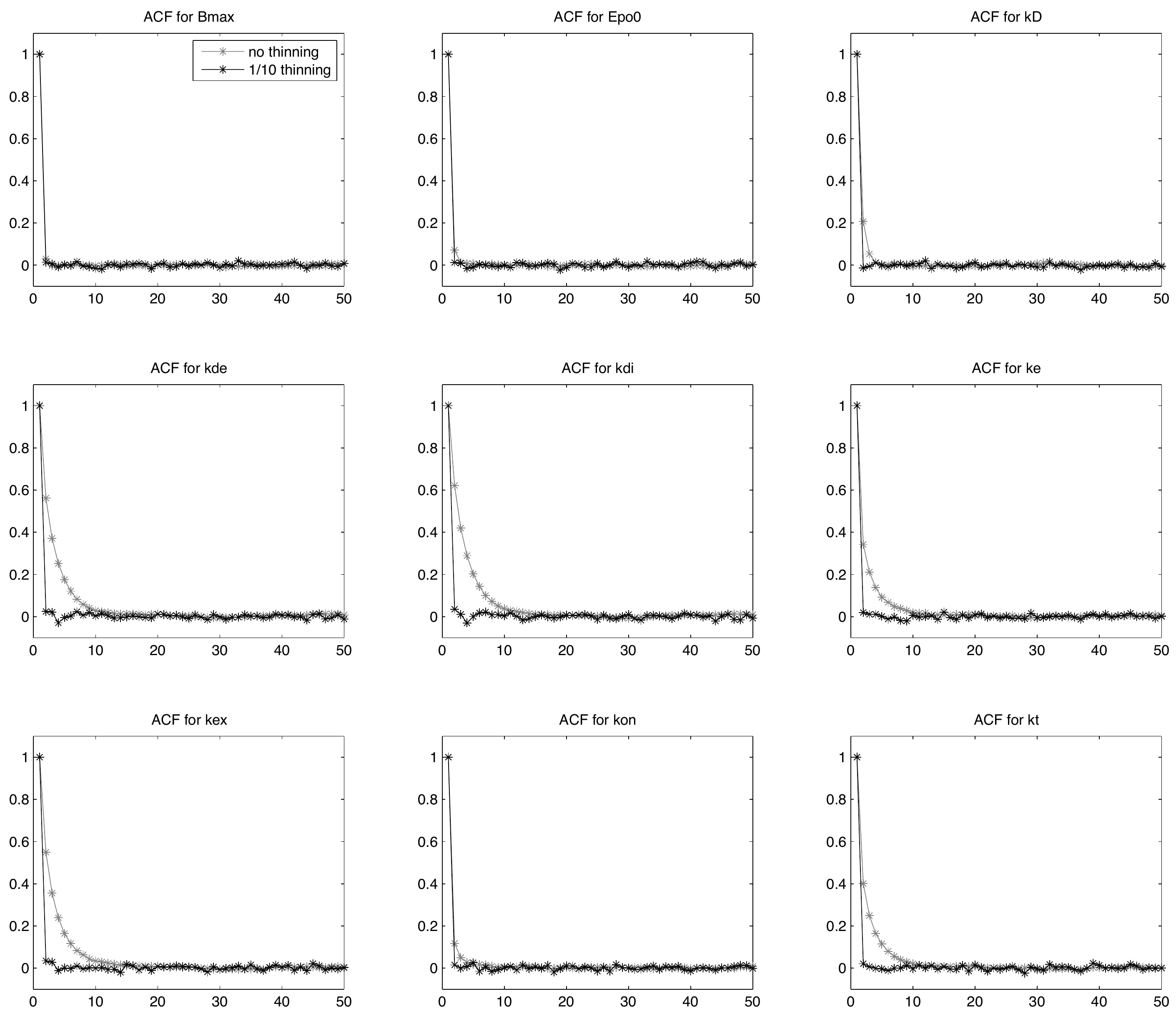}
\end{center}
\caption{ACF of the Markov chain using the MMALA algorithm for the extended setup: The figure displays the auto-correlation function (ACF) of the generated MCMC samples for each parameter along the horizontal axis. A thinning of 1/10 was used to increase independence of the samples. A rapidly decaying ACF indicates independence of the samples and convergence of the Markov chain.}
\label{acf2}
\end{figure}

\begin{figure}[b]
\begin{center}
\includegraphics[width=\textwidth]{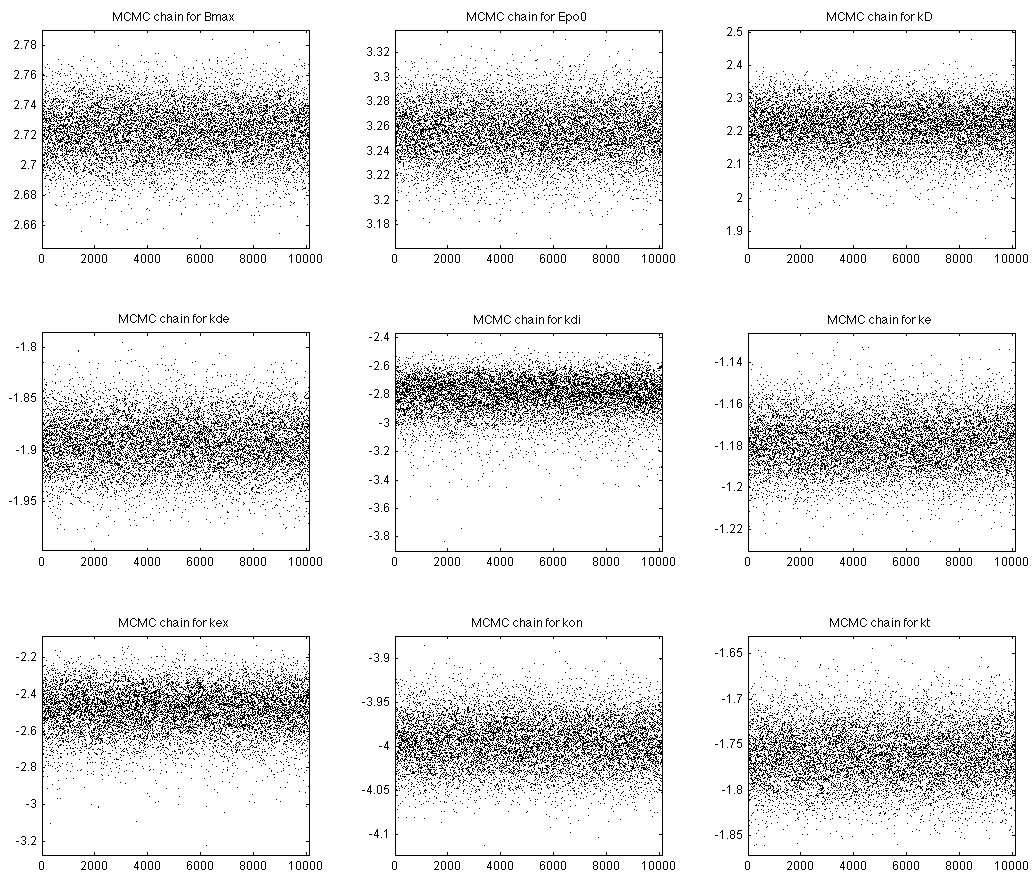}
\end{center}
\caption{Markov chains using the MMALA algorithm for the extended setup: The figure displays the generated samples for each parameter after thinning along the horizontal axis.}
\label{chains2}
\end{figure}

For the MCMC sampling, it is important to ensure that the Markov process starts in a high density region of $P(\theta|y)$ either by performing a burn-in or by searching for the MAP point, the latter option was used here. To monitor convergence of the Markov chain the generated samples can be checked for independence, e.g.~by computing their auto-correlation function (ACF), see figures \ref{acf1_naiv}, \ref{acf1}, \ref{acf2_naiv} and \ref{acf2}. If the samples are correlated thinning can be used to increase independence. Correlation in the samples can be visualised directly by the Markov chains, see figures \ref{chains1_naiv}, \ref{chains1}, \ref{chains2_naiv} and \ref{chains2}. Each sampling run was aimed at obtaining $10^4$ independent samples after applying thinning. For the SIM algorithm the 1/100 thinning still did not result in independent samples. The SIM algorithm turned out to be impractical both for the initial and extended setup. In contrast, the MMALA algorithm was much more efficient. A thinning of 1/100 and 1/10 for the initial and respectively the extended setup was sufficient to obtain $10^4$ independent samples. 

\begin{figure}[b]
\begin{center}
\includegraphics[width=\textwidth]{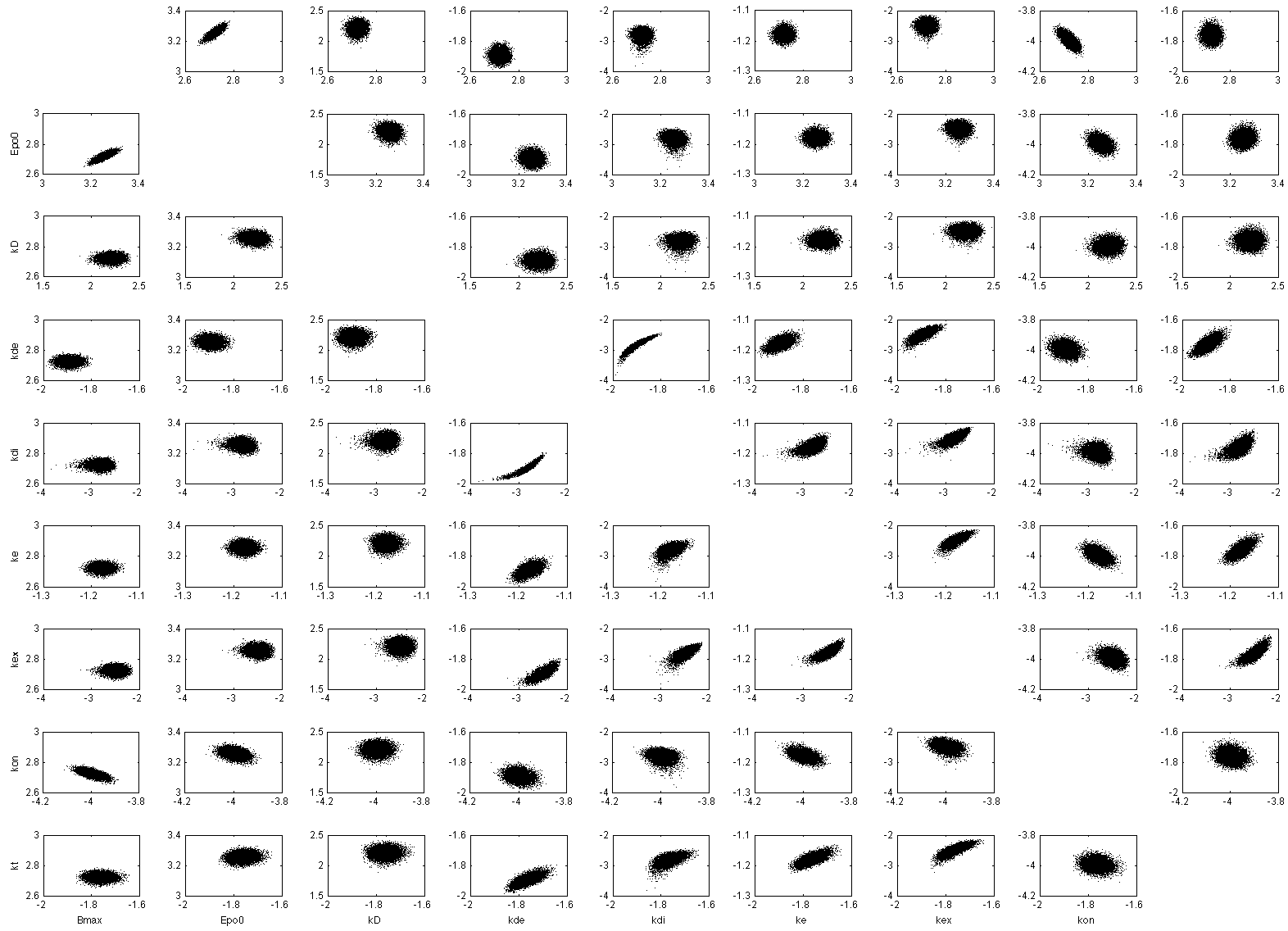}
\end{center}
\caption{Dependency structure of posterior samples: The results of the MCMC sampling allow for considering the full high dimensional posterior distribution. For example, non-linear relationship between parameter $k_{di}$ and $k_{de}$ are taken into account.}
\label{dep_mat}
\end{figure}

For the propagation of the parameter uncertainties contained in the posterior PDF to the prediction of the dynamics of unobserved molecular compounds, all trajectories corresponding to the samples shown in figure~\ref{chains2} or \ref{dep_mat} were evaluated. For each time point of the resulting set of curves a density estimate \citep{Azzalini:1997fk} was computed using the MATLAB function KSDENSITY. 

For the calculations MATLAB on a 2.1 GHz quad-core processor was used. The calculation of the profiles took $\approx5$ minutes for the initial setup and $\approx1$ minute of computation time for the extended setup. The generation of $10^6$ MCMC samples using the  MMALA algorithm took $\approx18.5$ hours for the initial setup and $\approx20$ minutes of computation time for the extended setup.

\bibliographystyle{unsrtnat}
\bibliography{template-suppl-extended.bbl}

\clearpage


\begin{table}
\begin{center}
\begin{tabular}{|r|rl|rl|rl|}
\hline
time / min & Epo\_ext  &/ cpm & Epo\_mem &/ cpm & Epo\_int &/ cpm \\
\hline
0 & 321194 & $\pm9461$ & - & & - &  \\
5 & 277333 &$\pm4984$ & 44698 & $\pm1731$ & 9125 &$\pm274$ \\ 
20 & 249910& $\pm3957$ & 44904 &$\pm1123$ & 50698 &$\pm346$ \\
60 & 231849 &$\pm12291$ & 27209 &$\pm2161$ & 83887 &$\pm2311$ \\ 
120 & 222047& $\pm2145$ & 23076 &$\pm1624$ & 102060 &$\pm4124$ \\ 
180 & 235203 &$\pm7075$ & 13680 &$\pm1057$ & 97952 &$\pm1493$ \\
240 & 251188 &$\pm6437$ & 10024 &$\pm1523$ & 82699 &$\pm4096$ \\
300 & 260945 &$\pm5200$ & 5447 &$\pm436$ & 65967 &$\pm3741$ \\
\hline
\end{tabular}
\caption{Experimental data: The experimentally accessible quantities are time-course measurements of radioactively labeled Epo collected in triplicates and units of counts per minute (cpm) over time in minutes (min) in different compartments: in the extracellular medium ($y_1$, Epo\_ext), on the cell membrane ($y_2$, Epo\_mem) and inside the cell ($y_3$, Epo\_int).}
\label{data}
\end{center}
\end{table}


\begin{table}
\begin{center}
\begin{tabular}{|l|l|l|}
\hline
parameter & $\log_{10}(\hat \theta)$ & unit \\
\hline
$B_{max}$ & $ +2.7126$ & pM \\
$Epo_\mathit{0}$ & $ +3.3075$ & pM \\
$k_D$ & $ +2.2148$ & pM \\
$k_{de}$ & $ -1.7850$ & 1/min \\
$k_{di}$ & $ -2.4977$ & 1/min \\
$k_e$ & $ -1.1259$ & 1/min \\
$k_{ex}$ & $ -2.0027$ & 1/min \\
$k_{on}$ & $ -3.9790$ & 1/(pM$\cdot$min) \\
$k_t$ & $ -1.4823$ & 1/min \\
$scale$ & $ +2.2311$ & cpm/pM \\
\hline
\end{tabular}
\caption{MAP estimates of the model parameters: The estimated values $\hat \theta$ were obtained by maximising the unnormalised posterior PDF by using the optimisation algorithm LSQNONLIN \citep{Coleman:1996fk} from MATLAB for the extended experimental setup.}
\label{map_pars}
\end{center}
\end{table}

\label{lastpage}